\documentclass[preprint,pre,longbibliography]{revtex4-1} 

\usepackage{natbib}
\usepackage{graphicx}
\usepackage{amssymb}
\usepackage{amsfonts}
\usepackage{amsmath}
\usepackage{amsbsy}
\usepackage{mathrsfs} 
\usepackage{color}
\usepackage{gensymb} 
\usepackage{hyperref} 
\usepackage{soul}%\st{...}
\usepackage{esint}
\usepackage{mymacros}

\newcommand{\mycite}[1]{\citeauthor{#1}~\cite{#1}}

\begin{document}

\title[Isotropically active particle closely fitting in a cylindrical channel]{Isotropically active particle closely fitting in a cylindrical channel: spontaneous motion at small P{\'e}clet numbers}

\author{ Rodolfo Brand{\~a}o}
\affiliation{Department of Mechanical and Aerospace Engineering, Princeton University, Princeton, New Jersey 08544, USA}

\begin{abstract}
Spontaneous motion due to symmetry breaking has been theoretically predicted for both active droplets and isotropically active particles in an unbounded fluid domain, provided their intrinsic P{\'e}clet number $\Pen$ exceeds a critical value. However, due to their inherently small $\Pen$, this phenomenon has yet to be experimentally observed for active particles. 
In this paper, we theoretically demonstrate that spontaneous motion for an active spherical particle closely fitting in a cylindrical channel is possible at arbitrarily small $\Pen$. Scaling arguments in the limit where the dimensionless clearance $\epsilon\ll1$ reveal that when $\Pen=O(\epsilon^{1/2})$, the confined particle reaches speeds comparable to those achieved in an unbounded fluid at moderate (supercritical) $\Pen$ values. We use matched asymptotic expansions in that distinguished limit, where the fluid domain decomposes into several asymptotic regions: a gap region, where the lubrication approximation applies; particle-scale regions, where the concentration is uniform; and far-field regions, where solute transport is one-dimensional. We derive an asymptotic formula for the particle speed, which is a monotonically decreasing function of $\overline{\Pen}=\Pen/\epsilon^{1/2}$ and approaches a finite limit as $\overline{\Pen}\searrow0$.  Our results could pave the way for experimental realisations of symmetry-breaking spontaneous motion in active particles.
\end{abstract}

\maketitle

\section{Introduction}\label{sec:intro}

Chemically active particles can undergo self-diffusiophoresis in liquid solutions by engendering concentration gradients through chemical reactions \citep{Paxton:04, Moran:17}. In their influential work, \mycite{Golestanian:07} proposed the first macroscale model for self-diffusiophoresis. It assumes Stokes flow conditions and diffusive transport of solute. On the particle boundary, chemical reactions are represented by a prescribed distribution of solute flux, while mechanical interactions with solute molecules are modelled by a diffusio-osmotic slip \citep{Anderson:89}. Due to the linearity of the governing equations and boundary conditions, self-propulsion is only predicted for particles exhibiting asymmetry in their shapes or physicochemical properties.

A nonlinear extension of that model, incorporating advective transport of solute, was later proposed by \mycite{Michelin:13}. The extended model predicts that a spherical particle with homogeneous physicochemical properties can self-propel via a symmetry-breaking mechanism. Two necessary conditions are: (i) the particle activity and slip coefficient must have the same sign; and (ii) the intrinsic P{\'e}clet number $\Pen$, indicating the strength of advection relative to diffusion, must be greater than $4$. While particles typically do not satisfy the latter condition \citep{Moran:17}, active droplets were experimentally observed to move spontaneously as a result of an analogous symmetry-breaking mechanism \citep{izri:14}. 

The macroscale description of \mycite{Michelin:13} has since served as the basis for multiple subsequent analyses of both asymmetry-driven \citep{Michelin:14, yariva:16, yariv:17, yariva:17} and spontaneous \citep{Saha:21, Desai:21, Desai:22, Picella:22,Schnitzer:22,Peng:23} motion, and has also been adopted as a reference model for active droplets \citep{michelin:23}. In particular, several recent extensions of \mycite{Michelin:13} have investigated the dynamics of isotropically active spheres constrained by solid boundaries \citep{yariva:16, yariv:17, Desai:21, Desai:22, Picella:22}. In purely hydrodynamic settings, the presence of boundaries naturally leads to increased viscous resistance to particle motion \citep{Happel:book}. Nonetheless, it has been shown that confinement promotes the spontaneous motion of active particles, leading to symmetry breaking at smaller $\Pen$ \citep{Desai:21, Desai:22}. 

In a recent study,  \mycite{Picella:22} presented a particularly dramatic effect of confinement on the spontaneous motion of active particles. Those authors simulated an active spherical particle moving through a cylindrical channel, with confinement quantified by the ratio $\epsilon$ of the minimum particle-channel clearance to the particle radius. They reported a bifurcation from a stationary-symmetric state when $\Pen$ exceeds a critical threshold. The reported critical $\Pen$ is a function of $\epsilon$, say $\Pen_c(\epsilon)$, that rapidly decreases with decreasing $\epsilon$ ($\Pen_c\approx 4$ for large $\epsilon$ and $\Pen_c\approx0.1$ for $\epsilon\approx1$). Unfortunately,  \mycite{Picella:22} could not investigate closely-fitting particles $(\epsilon\ll1)$ as their numerical method is not accurate for small $\epsilon$; in fact, they could not accurately determine $\Pen_c $ even for moderately small  $\epsilon$ ($\epsilon\lesssim1$).

\mycite{Picella:22} supplemented their simulations with analytical arguments, including a steady-state integral solute balance  (equation (5.3) therein) in a reference frame moving with the particle, assuming an infinitely long channel. We draw attention to two direct consequences of that balance. First, the problem is ill-posed for $\Pen = 0$, indicating that the small-$\Pen$ limit is singular. Second, any solution for $\Pen>0$ must involve particle motion at a non-zero speed relative to the channel walls, accompanied by fore-aft asymmetric flow and solute concentration. Consequently, there cannot be a bifurcation at finite $\Pen$ from a symmetric solution with a stationary particle to an asymmetric solution involving spontaneous motion. This latter conclusion, seemingly overlooked by \mycite{Picella:22}, reveals that the reported bifurcations can only be interpreted as numerical artefacts.

While the consequent numerical results in \mycite{Picella:22} may be questionable, their model---which is a direct extension of that in \mycite{Michelin:13}---provides a useful starting point for the analysis of confined active particles. Moreover, their simulations suggest that achieving spontaneous motion at small values of $\Pen$, which is representative of typical active particles, is possible under confinement.  Accordingly, a systematic investigation of the spontaneous motion of confined active particles at small $\Pen$ could pave the way for experimental demonstrations of this phenomenon in particles.

The present paper accordingly aims to analyse the motion of an isotropically active particle at small P{\'e}clet numbers ($\Pen\ll1$). As in \mycite{Picella:22}, we shall adopt the macroscale description of \mycite{Michelin:13}. We shall focus on closely fitting particles ($\epsilon\ll1$), corresponding to a regime where a detailed asymptotic description of the problem can be systematically constructed following the method of matched asymptotic expansions \citep{Hinch:book}---as in classical studies of moving particles closely fitting in channels \citep[see e.g.][]{Bungay:73}. We note that this regime has not been considered by \mycite{Picella:22} due to the aforementioned limitations of their numerical method.

The paper is structured as follows. We describe the physical problem in \S\ref{sec: pf} and formulate the mathematical problem in \S\ref{sec: dimensionless}. In \S\ref{sec: integral}, we derive integral balances for the mass and solute concentration. In \S\ref{sec: scales}, we perform a scaling analysis in the small-$\epsilon$ and small-$\Pen$ regime, which will allow us to determine the appropriate distinguished limit for the problem. In \S\ref{sec: anal}, we perform an asymptotic analysis of the problem in this distinguished limit. We discuss our results in \S\ref{sec: discuss}.

\section{Physical problem}\label{sec: pf}

\begin{figure}
\begin{center}
\includegraphics[scale=0.8]{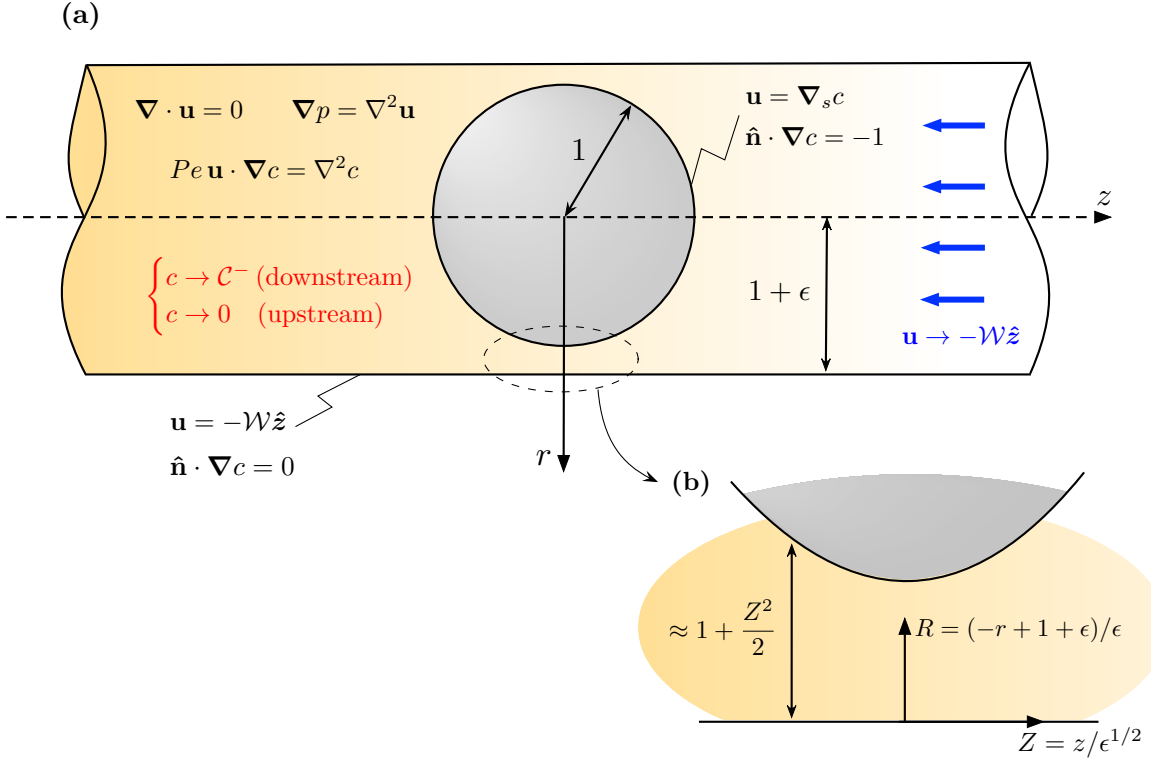}
\caption{ Schematic diagrams depicting (a) the dimensionless problem in the co-moving reference frame and (b) geometry of the gap region.}
\label{fig:sketch}
\end{center}
\end{figure}

Consider a chemically active spherical particle (radius $a^*$) moving within a cylindrical channel (radius $b^*>a^*$) filled with an otherwise quiescent fluid (viscosity $\mu^*$) that contains a single species of solute molecules (diffusivity $D^*$).

The particle exchanges solute with the fluid at a uniform rate per unit area $A^*$. This activity generates variations in the solute concentration, which are transported across the channel via advection and diffusion. Concurrently, changes in the concentration along the particle surface induce an effective diffusio-osmotic slip directly proportional to the surface gradient of the concentration. The proportionality constant, namely the slip coefficient $M^*$, is also assumed uniform.

We shall henceforth look for axisymmetric solutions of the problem where the particle moves along the channel axis at a constant speed. Since the physicochemical properties of the spherical particle are homogenous, the problem lacks a preferred direction along the channel axis.  Based on previous analyses \citep{Michelin:13, Picella:22}, we anticipate that symmetry breaking leading to spontaneous motion can occur if  $M^*$ and $A^*$ have the same sign. Motivated by that, we shall assume $M^* A^*>0$.

If the particle indeed exhibits symmetry-breaking spontaneous motion, dimensional arguments show that
\begin{equation}\label{eq: particle speed}
\text{particle speed} = \mathcal{W}\times \frac{M^*A^*}{D^*}.
\end{equation}
Here, $\mathcal{W}$ is the dimensionless speed, which depends upon two strictly-positive parameters, namely the dimensionless clearance,
\begin{equation}\label{eq:clear}
\epsilon = \frac{b^*-a^*}{a^*},
\end{equation}
and the intrinsic Péclet number,
\begin{equation}\label{eq:Peclet}
\Pen = \frac{M^*A^ *a^*}{D^{*2}}.
\end{equation}

It is convenient to discuss solute transport in the co-moving reference frame, where the flow and concentration fields are steady.  Far upstream, the solute concentration approaches a uniform value, which by causality must be independent of the particle's activity; far downstream, the concentration saturates at a different uniform value due to the accumulation of solute generated by the particle and advected by the flow.  Dimensional arguments show that, relative to its value far upstream,
\begin{equation}\label{eq: conc scale}
\text{solute concentration far downstream} = \mathcal{C}^-\times \frac{A^* a^*}{D^*}.
\end{equation}
Just like $\mathcal{W}$, the dimensionless downstream concentration $\mathcal{C}^-$ depends upon $\epsilon$ and $\Pen$.

\section{Dimensionless formulation}\label{sec: dimensionless}

We proceed to formulate the dimensionless steady-state problem in the co-moving reference frame, as illustrated in figure \ref{fig:sketch}(a).

We normalise lengths by $a^*$ and introduce the cylindrical coordinates $(r,\phi,z)$ with the origin at the centre of the particle; $r$ measures the radial distance from the channel axis, $\phi$ is the azimuthal angle, and the $z$-axis coincides with the channel axis and points in the direction of the particle's motion relative to the channel. The surface of the particle, denoted by $\mathcal{S}$, is parameterised as $r = f(z)$ with 
\begin{equation}
\qquad f(z) = \sqrt{1 - z^2}\quad\text{for}\quad|z|\leq 1
\end{equation}
and the channel wall is located at $r = 1 + \epsilon$. The fluid domain, denoted by $\mathcal{F}$, is given by $f(z) < r < 1 + \epsilon$ for $|z|\leq 1$ and $r < 1 + \epsilon$ for $|z|> 1$. 

We denote the excess solute concentration (normalised by $A^* a^*/D^*$), relative to its upstream value, by $c$; the pressure (normalised by $\mu^*M^*A^*/a^*D^*$) by $p$;  and the velocity (normalised by $M^*A^*/D^*$) by $\bu$. Axial symmetry dictates that the flow and concentration fields are independent of $\phi$ and that the velocity field can be written as 
\begin{equation}
\bu = u\boldsymbol{\hat{r}} + w\boldsymbol{\hat{z}},
\end{equation}
where $\boldsymbol{\hat{r}}$ and $\boldsymbol{\hat{z}}$ are unit vectors associated with the $r$ and $z$ coordinates, respectively.

The flow is governed by the Stokes equations,
\refstepcounter{equation}
$$
\label{eq: Stokes dl}
\bnabla p = \nabla^2 \bu,\qquad\bnabla\boldsymbol{\cdot} \bu =0 \quad\text{in}\quad \mathcal{F};
\eqno{(\theequation\mathrm{a},\!\mathrm{b})}
$$
the diffusio-osmotic slip condition,
\begin{equation}\label{eq: slip dl}
\bu = \bnabla_s c \quad \text{on}\quad  \mathcal{S},
\end{equation}
wherein $\bnabla_s = (\tI -\bn\bn)\boldsymbol{\cdot}  \bnabla$, with $\tI$ being the idemfactor and $\bn$ the outward unit normal;
the no-slip condition,
\begin{equation}\label{eq: noslip dl}
\bu = -\mathcal{W}\boldsymbol{\hat{z}} \quad \text{at}\quad r = 1+\epsilon;
\end{equation}
the far-field condition, specifying that the fluid is quiescent far from the particle in the laboratory frame,
\begin{equation}\label{eq: U far dl}
\bu \to -\mathcal{W}\boldsymbol{\hat{z}} \quad \text{as} \quad z\to\pm\infty;
\end{equation}
and the force-free condition,
\begin{equation} \label{eq:force-free dl}
\int_{\mathcal{S}} \bn \boldsymbol{\cdot} \left[ -p\tI + \left(\bnabla\bu\right) + \left(\bnabla\bu\right)^\dagger \right]\boldsymbol{\cdot}  \boldsymbol{\hat{z}}\, dS =0,
\end{equation}
with $\dagger$ denoting the tensor transpose and $dS$ denoting a differential area element.

The concentration field satisfies the advection--diffusion equation, 
\begin{equation}\label{eq:adv-dif dl}
 \Pen\, \bu\boldsymbol{\cdot} \bnabla c =\nabla^{2} c \quad\text{in}\quad \mathcal{F},
\end{equation}
subject to the activity condition, 
\begin{equation}\label{eq:act dl}
\bn\boldsymbol{\cdot}  \bnabla c = -1\quad \text{on}\quad \mathcal{S};
\end{equation}
the no-flux condition,
\begin{equation}\label{eq:noflux dl}
\bn\boldsymbol{\cdot}  \bnabla c = 0\quad \text{at}\quad r =1 + \epsilon;
\end{equation}
and the upstream and downstream conditions, respectively,
\refstepcounter{equation}
$$\label{eq:far-field c dl}
c\to 0 \quad\text{as}\quad z\to\infty,\qquad c\to \mathcal{C}^{-}\quad\text{as}\quad z\to-\infty.
\eqno{(\theequation\mathrm{a},\!\mathrm{b})}
$$

\section{Integral balances}\label{sec: integral}

We now derive exact integral balances that will facilitate our analysis.

To begin, we integrate the advection--diffusion equation \eqref{eq:adv-dif dl} over the entire fluid domain and apply the divergence theorem together with (\ref{eq: Stokes dl}b)--\eqref{eq: U far dl} and \eqref{eq:act dl}--\eqref{eq:far-field c dl}. This furnishes the relation 
\begin{equation}\label{eq: conc cons 2}
\Pen\, \mathcal{W}\mathcal{C}^-\left(1 + \epsilon\right)^2 = 4,
\end{equation}
which has been previously derived by \mycite{Picella:22}---equation (5.3) therein. This relation represents a balance between the net advective flux far downstream and the activity  \eqref{eq:act dl} integrated over the particle surface; in particular, it allows us to determine $\mathcal{C}^-$ once $\mathcal{W}$ is known. Clearly, \eqref{eq: conc cons 2} implies that the problem is ill-posed for $\Pen=0$ and that stationary solutions ($\mathcal{W}=0$) are impossible.

Next, we derive cross-sectional balances describing conservation of mass and solute between a cross-section of the channel located far upstream ($z\to\infty$) and another cross-section arbitrarily located ($z = \mathcal{Z}$) in the gap between the particle and the channel wall ($|\mathcal{Z}|\leq1$). To obtain the cross-sectional mass balance, we integrate the continuity equation (\ref{eq: Stokes dl}b) over the domain $\mathcal{F}_\mathcal{Z} = \{(r,z)\in\mathcal{F}\, | \,z\geq\mathcal{Z}\}$ and apply the divergence theorem in conjunction with \eqref{eq: slip dl}--\eqref{eq: U far dl}. This yields
\begin{equation}\label{eq: flux cons}
\int_{f(\mathcal{Z})}^{1 + \epsilon}w|_{z=\mathcal{Z}}r \, dr = -\frac{1}{2}\mathcal{W}\left(1 + \epsilon\right)^2.
\end{equation}

To obtain the cross-sectional solute balance, we integrate \eqref{eq:adv-dif dl} over $\mathcal{F}_\mathcal{Z}$ and apply the divergence theorem in conjunction with (\ref{eq: Stokes dl}b)--\eqref{eq: U far dl} and \eqref{eq:act dl}--\eqref{eq:far-field c dl}. This yields
\begin{equation}\label{eq: conc cons}
\int_{f(\mathcal{Z})}^{1 + \epsilon} \left(\Pen \,w c - \frac{\partial c}{\partial z} \right)\bigg|_{z=\mathcal{Z}} r \, dr = -1 + \mathcal{Z}.
\end{equation}
Note that the right-hand side represents the integration of the activity \eqref{eq:act dl} over the spherical cap $\mathcal{S}\cap\partial\mathcal{F}_\mathcal{Z}$.

\section{Closely fitting particle: scalings}\label{sec: scales}

We focus hereafter upon the narrow gap and small P{\'e}clet number regime, $\epsilon\ll1$ and $\Pen\ll1$. We begin with a scaling analysis. In addition to providing insight into the problem, our scalings will allow us to determine the appropriate distinguished limit for the subsequent asymptotic analysis.

Let us first describe the local geometry of the gap between the particle and channel, illustrated in figure \ref{fig:sketch}(b). There, the surface of the particle is approximately paraboloidal, and the particle-channel separation $h(z) =1 + \epsilon - f(z)$ can be expanded as
\begin{equation}\label{eq: pc sep} 
h(z)=\epsilon + \frac{1}{2}z^2 + \cdots \quad \text{as}\quad z\searrow0.
\end{equation}
Examining the first two terms of this expansion, we see that the separation remains of order $\epsilon$ over axial distances of order $\epsilon^{1/2}$. We postulate that the characteristic radial and axial lengthscales in the vicinity of the gap are given by $\epsilon$ and $\epsilon^{1/2}$, respectively.

Regarding the flow in the gap, the cross-sectional mass balance \eqref{eq: flux cons} implies that the axial velocity in the far-field, of order $\mathcal{W}$, is amplified by a factor of $\epsilon^{-1}$ as the narrow gap is approached. Therefore, the axial velocity in the gap scales as $\mathcal{W}/\epsilon$. The continuity equation (\ref{eq: Stokes dl}b) suggests that the radial velocity in the gap scales as $\mathcal{W}/\epsilon^{1/2}$; the axial component of the momentum equation (\ref{eq: Stokes dl}a) suggests that the pressure in the gap scales as $\mathcal{W}/\epsilon^{5/2}$; and the diffusio-osmotic slip condition \eqref{eq: slip dl} suggests that the solute concentration scales as $\mathcal{W}/\epsilon^{1/2}$.

Given the above, we now consider solute transport in the gap. With the advective flux $\Pen\,wc$ scaling as $\Pen\, \mathcal{W}^2/\epsilon^{3/2}$ and the diffusive flux $-\partial c/\partial z$ scaling as $\mathcal{W}/\epsilon$, the  cross-sectional solute balance \eqref{eq: conc cons} leads to the scaling relation
\begin{equation}\label{eq:scal rel}
\left(\frac{\Pen\, \mathcal{W}^2}{\epsilon^{3/2}}+\frac{\mathcal{W}}{\epsilon} \right)\times \epsilon \sim 1.
\end{equation}
This relation indicates that $\mathcal{W}$ is of order unity for $\Pen$ at most of order $\epsilon^{1/2}$ and asymptotically small for $\Pen\gg\epsilon^{1/2}$. (Curiously, our scalings arguments also suggest that order-unity speeds could be possible at small $\Pen$ even for $\epsilon$ of order unity. We return to this point in \S\ref{sec: discuss}.)

% Thus, if spontaneous motion is possible at small $\Pen$,  our scalings indicate that the particle can attain order unity speeds.

%we observe indications of spontaneous motion being possible and enhanced when both $\epsilon$ and $\Pen$ are small.

In light of our scaling arguments, the appropriate distinguished limit to be considered in the subsequent asymptotic analysis is
\begin{equation}\label{eq: Pen scale}
\Pen= O\!\left(\epsilon^{1/2}\right),
\end{equation}
which, by \eqref{eq:scal rel} corresponds to a scenario where advection and diffusion are comparable in the gap. In this distinguished limit, \eqref{eq:scal rel} implies that
\begin{equation}\label{eq: vel scale}
\mathcal{W}= O\!\left(1\right).
\end{equation}
Moreover, from \eqref{eq: conc cons 2}  we  obtain
\begin{equation}\label{eq: C scale}
\mathcal{C}^- = O\!\left(\epsilon^{-1/2}\right).
\end{equation}

\section{Closely fitting particle: asymptotic analysis}\label{sec: anal}

We now analyse the distinguished limit \eqref{eq: Pen scale}, represented by
\begin{equation}\label{eq: small-Pe}
\epsilon\ll1,\quad\Pen= \epsilon^{1/2} \overline{\Pen}
\end{equation}
with $\overline{\Pen}$ held fixed. In view of \eqref{eq: vel scale} and \eqref{eq: C scale}, we pose the expansions
\refstepcounter{equation}
$$
\label{eq: c u exp smallpe}
\mathcal{W} = \mathcal{W}_{0} + \cdots,\qquad\mathcal{C}^- = \epsilon^{-1/2} \mathcal{C}^-_{-1/2} + \cdots,
\eqno{(\theequation\mathrm{a},\!\mathrm{b})}
$$
where the leading-order balance of \eqref{eq: conc cons 2} reads
\begin{equation}\label{eq: conc cons 3}
\overline{\Pen}\,\mathcal{W}_0\mathcal{C}^-_{-1/2} = 4.
\end{equation}

\subsection{Remote regions}\label{ssec: ext and far}

We commence our asymptotic analysis by examining the flow and solute transport in the upstream ($z>0$) and downstream ($z<0$) remote regions of the channel. In those regions, advection and diffusion balance  \eqref{eq:adv-dif dl} over a  (dimensionless) lengthscale of order $\left(\Pen\,\mathcal{W}\right)^{-1}$, which, by \eqref{eq: Pen scale} and \eqref{eq: vel scale}, is equivalent to $\epsilon^{-1/2}$. Accordingly, we introduce the strained axial coordinate
\begin{equation}\label{eq: strain}
\tilde{z}= \epsilon^{1/2} z.
\end{equation}

Given that $\mathcal{W}$ is of order unity, the far-field condition \eqref{eq: U far dl} shows that the velocity in the remote regions is also of order unity. Moreover, when written in terms of $\tilde{z}$, the Stokes equations \eqref{eq: Stokes dl} together with \eqref{eq: noslip dl} and \eqref{eq: U far dl} demonstrate that this order-unity flow is uniform. It then follows from \eqref{eq: U far dl} that the velocity possesses the expansion
\begin{equation}
\bu=-\mathcal{W}_0\boldsymbol{\hat{z}} + \cdots.
\end{equation}

With a uniform velocity, it is evident that the leading-order concentration must be a function of $\tilde{z}$ alone. To analyse solute transport, we separately examine the upstream and downstream regions.  In the upstream region, we assume, based on \eqref{eq: C scale}, that the solute concentration is of order $\epsilon^{-1/2}$. This motivates the expansion
\begin{equation}\label{eq: c}
c = \epsilon^{-1/2}\tilde{c}^{+}_{-1/2} + \cdots.
\end{equation}
Expanding \eqref{eq:adv-dif dl} and (\ref{eq:far-field c dl}a), we find that $\tilde{c}^{+}_{-1/2}$ satisfies the one-dimensional advection--diffusion equation,
 \begin{equation}\label{eq: adv diff far}
-\overline{\Pen}\,\mathcal{W}_0 \frac{d \tilde{c}^{+}_{-1/2} }{d\tilde{z}} = \frac{d^2 \tilde{c}^{+}_{-1/2} }{d\tilde{z}^2}, 
\end{equation}
subject to the far-field condition (cf.\,(\ref{eq:far-field c dl}a))
\begin{equation}\label{eq: c far far }
\tilde{c}^+_{-1/2} \to 0\quad \text{as}\quad \tilde{z}\to-\infty.
\end{equation}
The solution to the above problem is
\begin{equation}\label{eq: c resul 1}
\tilde{c}^+_{-1/2} = \mathcal{C}^+_{-1/2}\exp{\left(-\frac{\tilde{z}}{\overline{\Pen}\,\mathcal{W}_0}\right)},
\end{equation}
where the constant $\mathcal{C}^+_{-1/2}$ is yet to be determined.

In the downstream region, the concentration is also of order $\epsilon^{-1/2}$, in accordance to \eqref{eq: C scale}, so we pose the expansion 
\begin{equation}
c = \epsilon^{-1/2}\tilde{c}^-_{-1/2}+\cdots.
\end{equation}
The leading-order concentration $\tilde{c}^+_{-1/2}$ satisfies the same one-dimensional advection--diffusion equation as in \eqref{eq: adv diff far} but now subject to the condition $\tilde{c}^+\to\mathcal{C}^-_{-1/2}$ as $\tilde{z}\to\infty$ (cf. (\ref{eq:far-field c dl}b)). The solution is
\begin{equation}\label{eq: c resul 2}
\tilde{c}^-_{-1/2}= \mathcal{C}^-_{-1/2}.
\end{equation}

Accordingly, the leading-order concentration decays exponentially in the upstream remote region \eqref{eq: c resul 1} and is uniform in the downstream remote region \eqref{eq: c resul 2}.

\subsection{Particle-scale regions}\label{ssec: particle size}

In the downstream region adjacent to the particle, at distances of order unity and $z<0$, the concentration remains uniform to leading order and equal to \eqref{eq: c resul 2}. This is because the order-unity activity \eqref{eq:act dl} cannot locally affect the order-$\epsilon^{-1/2}$ concentration. (We also do not expect an asymptotically large solute flux emanating from the gap, as any such flux would imply a concentration in the gap $\gg \epsilon^{-1/2}$, contradicting our scalings.)

Similarly, the concentration is uniform to leading order in the vicinity of the particle with $z>0$. There, asymptotic matching with \eqref{eq: c resul 1} shows that, to leading order, $c$ is identically given by $\epsilon^{-1/2}\mathcal{C}^+_{-1/2}$.

Therefore, in the vicinity of the particle, at distances of order unity, we expand the concentration as
\begin{equation}
\label{gap: c particle scale smallpe}
c = \begin{cases}
\epsilon^{-1/2}\mathcal{C}^-_{-1/2}+\cdots,&\quad z<0,\\
\epsilon^{-1/2}\mathcal{C}^+_{-1/2}+\cdots,&\quad z>0.
\end{cases}
\end{equation}
As we shall see, the constants  $\mathcal{C}^-_{-1/2}$ and  $\mathcal{C}^+_{-1/2}$ are generally distinct.
%expand the solute concentration as

 \subsection{Gap region}

Following the scalings in \S\ref{sec: scales}, we introduce stretched coordinates in the gap region
\refstepcounter{equation}
$$\label{eq:coord}
R= (-r + 1 + \epsilon)/\epsilon,\quad  Z= z/\epsilon^{1/2}.
 \eqno{(\theequation\mathrm{a},\!\mathrm{b})}
$$
In terms of these, the particle-channel separation \eqref{eq: pc sep} can be written as
 \begin{equation}\label{eq: HZ}
R= H(Z) + O(\epsilon),\quad H(Z) = 1 + \frac{Z^2}{2}.
 \end{equation}
Moreover, we expand the solute concentration as 
\begin{equation}\label{eq: conc gap}
c = \epsilon^{-1/2}C_{-1/2}(R,Z) + \cdots;
\end{equation}
the pressure as
\begin{equation}\label{eq: P gap}
p = \epsilon^{-5/2}P_{-5/2}(R,Z) + \cdots;
\end{equation}
and the axial and radial velocity, respectively as
\refstepcounter{equation}
$$\label{eq: v gap}
 u = \epsilon^{-1/2}U_{-1/2}(R,Z) + \cdots,\quad  w = \epsilon^{-1}W_{-1}(R,Z)+ \cdots.
\eqno{(\theequation\mathrm{a},\!\mathrm{b})}
$$

Substituting \eqref{eq: c u exp smallpe} and \eqref{eq: 	HZ}--\eqref{eq: P gap} into the integral balances \eqref{eq: flux cons} and \eqref{eq: conc cons} and retaining leading-order terms, we obtain
\begin{subequations}
  \label{eq: int rel gap}
\begin{gather}
 \int_{0}^{H(Z)} W_{-1}\,dR=-\frac{1}{2}\mathcal{W}_0,\quad \nonumber \tag{\theequation a}\\ 
\int_{0}^{H(Z)}\left(\overline{\Pen} \,W_{-1} C_{-1/2} - \frac{\partial C_{-1/2}}{\partial Z}\right)\,dR= -1 \tag{\theequation  b},
\end{gather}
 \end{subequations}
which are valid for all $Z$.

\subsubsection{Solute transport}

Expanding \eqref{eq:adv-dif dl}--\eqref{eq:noflux dl} using \eqref{eq:coord}--\eqref{eq: P gap}, we find that the leading-order concentration satisfies
\begin{equation}\label{eq: c12 problem}
\frac{\partial^2 C_{-1/2}}{\partial R^2}=0,
\end{equation}
with boundary conditions
\begin{equation}\label{eq: c12 problem 2}
\frac{\partial C_{-1/2}}{\partial R}=0 \quad\text{at}\quad R=0 \quad\text{and}\quad R= H(Z).
\end{equation}
Thus, $C_{-1/2}$ is independent of $R$. We can therefore evaluate the integral in (\ref{eq: int rel gap}b) using (\ref{eq: int rel gap}a) to arrive at the transport equation
\begin{equation} \label{eq: transport c gap 2}
 H(Z)\frac{dC_{-1/2}}{dZ} +\frac{1}{2}\overline{\Pen}\,\mathcal{W}_0 C_{-1/2} = 1.
 \end{equation}
Moreover, asymptotic matching with \eqref{gap: c particle scale smallpe} for $z<0$ furnishes the far-field condition
  \begin{equation} \label{eq: c far 3}
C_{-1/2}\to\mathcal{C}_{-1/2}^-\quad \text{as}\quad Z\to-\infty.
 \end{equation} 

Solving \eqref{eq: transport c gap 2} subject to \eqref{eq: c far 3}, we find 
\begin{equation}\label{eq: C in gap}
C_{-1/2} = \frac{2}{\overline{\Pen}\, \mathcal{W}_0} + \left(\mathcal{C}_{-1/2}^-- \frac{2}{\overline{\Pen}\, \mathcal{W}_0} \right)\exp \left[-\frac{\overline{\Pen}\,\mathcal{W}_0}{\sqrt{2}} \left( \frac{\pi}{2}+ \text{arctan} \frac{Z}{\sqrt{2}}\right)\right],
\end{equation} 
which can be rewritten with the aid of \eqref{eq: conc cons 3} as 
\begin{equation}\label{eq: C in gap 2}
C_{-1/2} = \frac{2}{\overline{\Pen}\, \mathcal{W}_0}\left\{1 + \exp \left[-\frac{\overline{\Pen}\,\mathcal{W}_0}{\sqrt{2}} \left(\frac{\pi}{2}+ \text{arctan} \frac{Z}{\sqrt{2}}\right)\right]\right\},
\end{equation} 
For later reference, the derivative of $C_{-1/2}$ with respect to $Z$ is given by
\begin{equation}\label{eq: c gradient gap}
\frac{dC_{-1/2}}{dZ} = -\frac{1}{H(Z)} \exp \left[-\frac{\overline{\Pen}\,\mathcal{W}_0}{\sqrt{2}} \left(\frac{\pi}{2}+ \text{arctan} \frac{Z}{\sqrt{2}}\right)\right].
  \end{equation}

Taking $Z\to\infty$ in \eqref{eq: C in gap 2} shows that the concentration field approaches a uniform value at large distances in the gap.  Thus, asymptotic matching between \eqref{gap: c particle scale smallpe} and \eqref{eq: C in gap 2} for $z>0$ yields
  \begin{equation}
\tilde{\mathcal{C}}^+= \frac{2}{\overline{\Pen}\, \mathcal{W}_0}\left[1 + \exp \left(-\frac{\pi \overline{\Pen}\,\mathcal{W}_0}{\sqrt{2}}\right)\right].
\end{equation}
Note that $\mathcal{W}_0$ remains to be determined.
  
\subsubsection{Lubrication flow}

Expanding \eqref{eq: Stokes dl}--\eqref{eq: noslip dl} using \eqref{eq:coord}--\eqref{eq: P gap}, we find that the flow in the gap is governed by 
\refstepcounter{equation}
$$\label{eq: stokes 2}
\frac{\partial P_{-5/2}}{\partial R}=0,\qquad\frac{\partial^2 W_{-1}}{\partial R^2}= \frac{\partial P_{-5/2}}{\partial Z};
\eqno{(\theequation\mathrm{a},\!\mathrm{b})}
$$
the no-slip condition,
  \begin{equation}\label{eq: bcs lub 1}
W_{-1}= 0\quad\text{at}\quad R=0;
 \end{equation}
and the diffusio-osmotic condition,
  \begin{equation}\label{eq: bcs lub 2}
W_{-1}= \frac{d C_{-1/2}}{d Z} \quad \text{at}\quad R = H(Z).
 \end{equation}
  
With $P_{-5/2}$ is independent of $R$ (see (\ref{eq: stokes 2}a)), we can integrate equation (\ref{eq: stokes 2}b) using \eqref{eq: bcs lub 1} and \eqref{eq: bcs lub 2}. This furnishes an expression for the axial velocity in terms of the pressure and concentration gradients,
\begin{equation}\label{eq: U1}
W_{-1}= \frac{R\left(R - H\right)}{2}\frac{d P_{-5/2}}{dZ}+ \frac{R}{H}\frac{d C_{-1/2}}{d Z}.
\end{equation}
Next, integrating \eqref{eq: U1} from $R=0$ to $R = H(Z)$ using (\ref{eq: int rel gap}a) gives
 \begin{equation}
-\frac{H^3}{12}\frac{d P_{-5/2}}{dZ}+ \frac{H}{2}\frac{d C_{-1/2}}{d Z} = -\frac{1}{2}\mathcal{W}_0, 
 \end{equation}
 which can be rearranged as
 \begin{equation}\label{eq: P52}
\frac{d P_{-5/2}}{dZ} = \frac{6}{H^3}\mathcal{W}_0 + \frac{6}{H^2} \frac{d C_{-1/2}}{d Z}.
 \end{equation}

 \subsection{Particle speed}
 
 \begin{figure}
\begin{center}
\includegraphics[scale=0.35]{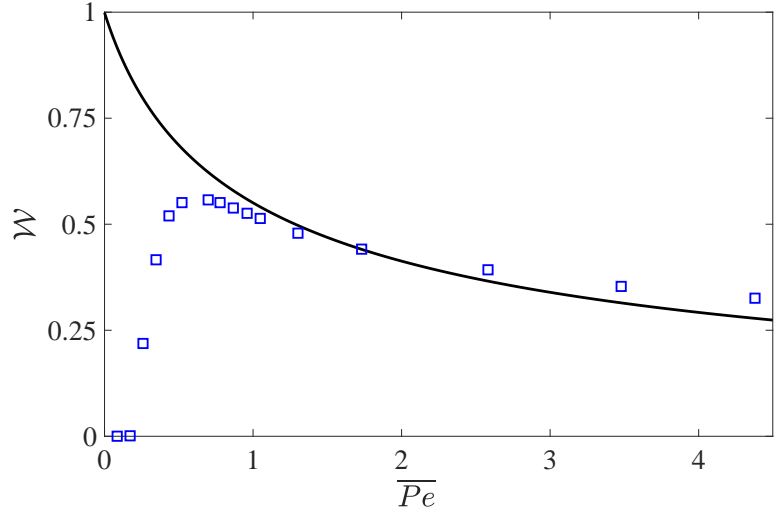}
\caption{Dimensionless particle speed $\mathcal{W}$ as a function of $\overline{\Pen}$. The solid line depicts the leading-order approximation calculated using \eqref{eq: U0 resul}. The blue squares represent the numerical data extracted from \mycite{Picella:22} for $\epsilon = 1/3$.}
\label{fig:comp}
\end{center}
\end{figure}

We can now determine the particle speed $\mathcal{W}_0$. First, we note that the pressure drop along the gap is given by 
\begin{equation}\label{eq: delta p}
\epsilon^{-5/2} \int_{-\infty}^{\infty}\frac{d P_{-5/2}}{dZ}\,dZ + \cdots.
 \end{equation}
This order-$\epsilon^{-5/2}$ pressure drop must vanish  \citep{Bungay:73}, 
  \begin{equation}\label{eq: delta p 2}
 \int_{-\infty}^{\infty}\frac{d P_{-5/2}}{dZ}\,dZ= 0;
 \end{equation}
 if this condition is not met, asymptotic matching with the particle-scale regions shows that the pressure difference between the right hemisphere ($z>0$) and left hemisphere ($z<0$) of the particle is of order $\epsilon^{-5/2}$, resulting in a net force of the same order, which contradicts the force-free condition \eqref{eq:force-free dl}.
    
Substituting \eqref{eq: P52} into \eqref{eq: delta p 2} then gives
 \begin{equation}
6\,\mathcal{W}_0\int_{-\infty}^{\infty} \frac{dZ}{H^3} + 6 \int_{-\infty}^{\infty} \frac{d C_{-1/2}}{d Z} \frac{dZ}{H^2}=0.
 \end{equation}
The integrals can be evaluated in closed form upon substitution of \eqref{eq: HZ} and \eqref{eq: c gradient gap}. This yields
\begin{equation}\label{eq: U0 resul}
 \mathcal{W}_0 = \frac{\sqrt{2}}{\pi}\frac{1 - \exp{\left(-\pi\overline{\Pen}\,\mathcal{W}_0/\sqrt{2}\right)}}{ \overline{\Pen}\,\mathcal{W}_0 + 5\overline{\Pen}^3\,\mathcal{W}_0^3/32 + \overline{\Pen}^5\,\mathcal{W}_0^5/256},
\end{equation}
which provides  $\mathcal{W}_0$ as a implicit function of $\overline{\Pen}$. The downstream concentration $\mathcal{C}^-_{-1/2}$ can then be determined from \eqref{eq: conc cons 3},
 \begin{equation}\label{eq: conc cons 4}
\mathcal{C}^-_{-1/2} = 2\pi\sqrt{2}\,\mathcal{W}_0 \frac{1+ 5\overline{\Pen}^2\,\mathcal{W}_0^2/32 + \overline{\Pen}^4\,\mathcal{W}_0^4/256}{1 - \exp{\left(-\pi\overline{\Pen}\,\mathcal{W}_0/\sqrt{2}\right)}}.
\end{equation}
In particular, we note the limits 
\refstepcounter{equation}
$$\label{eq: U0 limit}
 \mathcal{W}_0\to 1,\qquad \mathcal{C}^-_{-1/2}\sim \frac{4}{ \overline{\Pen}}\quad\text{as}\quad\overline{\Pen}\searrow0.
\eqno{(\theequation\mathrm{a},\!\mathrm{b})}
$$
 
\section{Discussion and future perspectives}\label{sec: discuss}

We have investigated the spontaneous motion of an isotropically active spherical particle closely fitting in a cylindrical channel ($\epsilon\ll1$) at small P{\'e}clet numbers ($\Pen\ll1$). Our focus has been on scenarios where the particle moves along the channel axis and the concentration and flow fields are steady in the co-moving reference frame. Despite spontaneous motion being impossible for a particle in an unbounded domain at small $\Pen$, our analysis has demonstrated its existence for a particle in a channel, showing that order-unity speeds are attained for  $\Pen$ of order $\epsilon^{1/2}$. By systematically analysing this distinguished limit, we have derived asymptotic approximations for the dimensionless particle speed $\mathcal{W}$ and downstream concentration $\mathcal{C}^-$, as well as for the flow and concentration fields along the channel. 

Figure \ref{fig:comp} depicts the leading-order particle speed \eqref{eq: U0 resul} as a function of the rescaled P{\'e}clet number $\overline{\Pen} = \Pen/\epsilon^{1/2}$. There is no bifurcation, in accordance with the integral balance \eqref{eq: conc cons 2}. Rather, a unique solution exists for all values of $\overline{\Pen}$, with $\mathcal{W}$  approaching unity as $\overline{\Pen}$ tends to zero (cf.\,(\ref{eq: U0 limit}a)) and decreasing monotonically with $\overline{\Pen}$.  As the problem is ill-posed for $\Pen=0$ (cf.\,\eqref{eq: conc cons 2}), the limiting value $\mathcal{W}=1$ cannot be attained in practice. Despite $\mathcal{W}$ remaining finite, other physical quantities diverge as $\overline{\Pen}$ tends to zero, such as $\mathcal{C}^-$ (\ref{eq: U0 limit}b) and the lengthscale $(\Pen\,\mathcal{W})^{-1}$ over which the concentration decays far upstream (see \eqref{eq: c resul 1}).  Accordingly, $\mathcal{W}=1$ provides an unattainable upper bound for the leading-order speed in the small-$\epsilon$ and small-$\Pen$ regime. We note that this upper bound also holds for all values of $\epsilon$ and $\Pen$ in the simulations of \mycite{Picella:22}.

In figure \ref{fig:comp}, we also compare our asymptotic results with the numerical data of \mycite{Picella:22} for $\epsilon=1/3$. Despite $\epsilon$ hardly being small, we observe a relatively good agreement that appears to improve as $\overline{\Pen}$ decreases until $\overline{\Pen}\approx1$. For smaller $\Pen$, numerics and asymptotic disagree, with the numerical speed decaying as $\overline{\Pen}$ decreases further and seemingly reaching zero at a small but finite $\Pen$ value. This numerical bifurcation violates \eqref{eq: conc cons 2}, and hence must constitute a numerical artefact, which explains the observed discrepancies between numerics and asymptotics.

The spurious numerical bifurcations in \mycite{Picella:22} might originate from how the computational domain is truncated at remote cross-sections. In detail, the simulations employed a long but finite channel, with both ends of the channels co-moving with the particle and the concentration field subject to homogenous Neumann conditions at the ends. Simulations in the truncated domain with said conditions should approximate the problem formulated herein for an infinite channel, provided the truncated channel is long relative to other lengthscales of the problem. In particular, in the small $\Pen$ regime studied herein, the largest lengthscale of the problem is the aforementioned  $(\Pen\,\mathcal{W})^{-1}$, which becomes arbitrarily large. Hence, numerical end effects inevitably introduce discrepancies between numerics and asymptotics. 

It is natural to ask whether previously unexplored physical mechanisms, as opposed to numerical artefacts, might also become important at sufficiently small $\Pen$. Incorporating such mechanisms might also regularise the problem, making it well-posed for $\Pen=0$.  A regularised physical problem could involve a particle in a realistic finite channel, such as a channel with open ends connected to large reservoirs.  For $\Pen=0$, the solute would leak to the reservoirs, and diffusion would cause the excess concentration to decay away from the channel, thus regularising the problem. To extend our analysis to finite channels, we could adopt a method analogous to that used by \mycite{Sherwood:21}, who considered the electrophoresis of a closely fitting sphere in a cylindrical channel with open ends. A systematic analysis of finite channels would likely yield results that differ from those in \mycite{Picella:22}, where the consideration of finite channels is merely a consequence of numerical truncation. Indeed, the effective end conditions employed by \mycite{Picella:22} do not represent real finite channels, first because they do not allow for any solute leakage, and second because the position of the channel ends should be fixed in the laboratory frame rather than in the co-moving frame.

Another physical mechanism that might regularise the problem is solute absorption due to a chemical reaction in the fluid \citep{De:13}. As advection gets weaker, we expect that bulk absorption, however small, could hinder the solute from penetrating further into the channel. Even in the case of a stationary particle, where a net advective flux is absent, absorption would still lead the excess concentration to decay far away from the particle, thus making the problem well-posed. Extending our analysis to include weak bulk absorption should be straightforward, as absorption would only be relevant far from the particle, leaving the analysis in the gap region intact.

Several other extensions of our analysis, not necessarily associated with regularisation mechanisms, are also worth pursuing. For example, it would be interesting to analyse a regime where $\epsilon$ is moderate and $\Pen$ is small. When $\epsilon$ increases from small to moderate values with $\Pen$ held fixed, \eqref{eq: U0 resul} shows that the speed grows and approaches unity. (This limiting process corresponds to $\overline{\Pen}\searrow0$, see \eqref{eq: small-Pe}.) Accordingly, although \eqref{eq: U0 resul} ceases to be valid for $\epsilon = O(1)$, it nonetheless suggests that order-unity speeds are attainable at arbitrarily small $\Pen$ for moderate values of $\epsilon$. An analysis of this moderate-$\epsilon$ regime could also bridge our small-$\epsilon$ solution to the corresponding solution for large $\epsilon$. Regarding the latter, we expect that the large-$\epsilon$ speed remains non-zero for all $\Pen$, in accordance with \eqref{eq: conc cons 2}, but becomes asymptotically small in $\epsilon^{-1}$ for $\Pen\leq 4$; for $\Pen>4$, we expect a stable solution with $\mathcal{W}$ of order unity, approaching the well-known solution for an active particle in an unbounded domain \citep{Michelin:13}. Other natural regimes to explore include small $\epsilon$ with moderate or large $\Pen$. Although the scalings presented in \S\ref{sec: scales} do not suggest other distinguished limits when $\epsilon\ll1$, a preliminary analysis for larger $\Pen$ indicates that a distinguished limit emerges when $\Pen$ is of order unity.

An analysis of the distinguished limit $\epsilon\ll1$ with $\Pen=O(1)$ might also provide the first steps towards extending our theory to active droplets, which typically exhibit moderate to large values of $\Pen$. Of course, a proper analysis of spontaneous motion in droplets must account for the internal flow within the droplet, as well as Marangoni stresses \citep{michelin:23}. In addition, a droplet substantially deforms inside a channel and becomes nearly cylindrical as its volume increases; hence a closely-fitting droplet is fundamentally different from the active spherical particles considered here. We note that the spontaneous motion of active droplets with large $\Pen$ has been recently demonstrated by \mycite{de:21}, who observed that droplets attain large speeds in the channel relative to unbounded domains. Those authors rationalised these findings using an intuitive analysis based on an analogy with the classical Bretherton \mycite{Bretherton:61} problem. It would be worthwhile to revisit that analysis following a systematic approach using matched asymptotic expansions.

Lastly, we note that the focus on active droplets in the literature is partly due to a commonly held view that spontaneous motion cannot occur for particles due to their small $\Pen$ \citep{michelin:23}.   However, our results challenge this view and show that the spontaneous motion of a strongly confined particle in a channel is possible at small $\Pen$. Thus, our theory has direct implications for the study of active particles and, in particular, could inform experimental realisations of symmetry-breaking spontaneous motion.

\bibliography{refs}
\end{document}